\documentclass[preprint]{aastex}

\usepackage{rotating}
\shorttitle{Tilt, Warp, and Precessing Disks}
\shortauthors{Montgomery}

\begin{document}

\title{Tilt, Warp, and Simultaneous Precessions in Disks}
\author{M.M. Montgomery\altaffilmark{1} }
\affil{$^{1}$Department of Physics, University of Central Florida, Orlando, FL  32816, USA}

\begin{abstract}
Warps are suspected in disks around massive compact objects.  However, the proposed warping source -- non-axisymmetric radiation pressure -- does not apply to white dwarfs.  In this letter we report the first Smoothed Particle Hydrodynamic simulations of accretion disks in SU UMa-type systems that naturally tilt, warp, and simultaneously precess in the prograde and retrograde directions using white dwarf V344 Lyrae in the $\emph{Kepler}$ field as our model. After $\sim$79 days in V344 Lyrae, the disk angular momentum $\bf{\emph{L}_{d}}$ becomes misaligned to the orbital angular momentum $\bf{\emph{L}_{o}}$.  As the gas stream remains normal to $\bf{\emph{L}_{o}}$, hydrodynamics (e.g., the lift force) is a likely source to disk tilt. In addition to tilt, the outer disk annuli cyclically change shape from circular to highly eccentric due to tidal torques by the secondary star.  The effect of simultaneous prograde and retrograde precession is a warp of the colder, denser midplane as seen along the disk rim.  The simulated rate of apsidal advance to nodal regression per orbit nearly matches the observed ratio in V344 Lyrae.  
\end{abstract}


\section{Introduction}
V344 Lyrae (Wood et al. 2011) is a short-period SU UMa-type of the Dwarf Novae (DN) subclass to the photometrically active Cataclysmic Variables (CVs).  SU UMa's consist of a late-type main-sequence star that fills its Roche lobe and transfers its gaseous material to an accretion disk around a white dwarf.  In addition to outbursts (e.g., Lasota 2001), which are periodic increases in brightness due to a thermal instability in the disk, SU UMa systems exhibit superoutbursts that last a few times longer and are up to a magnitude brighter.  The source to superoutbursts is a tidal instability:  The disk expands outward until enough mass has expanded beyond the $m$=3 eccentric inner Lindblad resonance radius, at which time the outer disk annuli are tidally forced into cyclically changing shape from circular to highly eccentric.

A characteristic of superoutbursts is large-amplitude positive superhumps that have a disk origin (Horne 1984).  In addition to these positive superhumps, negative superhumps are also sometimes found in the same system.  Like positive superhumps, negative superhumps also have a disk origin (Montgomery 2012a).  Systems that show positive or negative superhumps in light curves also sometimes show prograde or retrograde precession, respectively, suggesting superhumps and precession are intrinsically linked. Numerical simulations show disks cyclically changing shape from circular to highly eccentric, producing a saturated positive $\emph{early}$ superhump that has a slightly longer period than the orbital period.  Tidal torques on the asymmetric density disk result in disk prograde precession; the secondary needs to advance slightly more than one orbital period to meet up with the next shape-change cycle, and several cycles are needed for each prograde precession. Two asymmetric spiral-like density waves in the outer-to-mid annuli advance $\sim$180$^{o}$ in the prograde direction in one positive superhump period as shown in Simpson \& Wood (1998), for example, and may be induced when higher order harmonics are present (Montgomery 2012b).  Other numerical simulations show disks tilting, producing a negative superhump that has a slightly shorter period than the orbital period (Montgomery 2012a).  Tidal torques on the tilted disk result in disk retrograde precession (Montgomery 2009a); the secondary meets up with the second node slightly before one orbital period, and several cycles are needed for each retrograde precession (Montgomery 2012a).  For CV DN SU UMa-type systems and their higher-mass transfer rate analogs, nova-likes (NLs), both precessions are seen either in series or simultaneously.  As noted in Skillman et al. (1998) for TT Ari, understanding how a differentially rotating fluid disk can manage $\emph{one}$ well-defined precession frequency is difficult in itself, let alone two. To date, both precessions have not yet been shown to occur naturally in a numerical simulation, although some artificially induce (Wood et al. 2011). 

In this work, we show how a differentially rotating fluid disk can simultaneously precess in both directions.  We use the same 25,000 particle SPH code (Simpson \& Wood, 1998) that has successfully produced either positive superhumps and prograde precession (e.g., Simpson \& Wood, 1998) or negative superhumps and retrograde precession (Montgomery 2012a).  We use V344 Lyrae, an SU UMa system in the $\emph{Kepler}$ field, as our model. V344 Lyrae experiences outbursts and superoutbursts; has $\sim$0.25 mag positive and $\sim$0.8 mag negative superhumps with periods 2.20 and 2.06 hours, respectively; and has an orbital period 2.11 hours (Wood et al. 2011).  

In \S2, we briefly discuss the SPH code, and in \S3, we present the simulation results.  In \S4, we discuss the results, and in \S5, we provide conclusions and future work.

\section{3D SPH Numerical Code}
The 3D SPH codes used and described in Wood et al. (2000), Montgomery (2004, 2009a), Wood et al. (2009), and Wood et al. (2011) have their roots in Simpson (1995), a code that applies to non-magnetic CV systems.  These later versions of the root code have e.g., more particles (i.e., more than 100,000), modified particle shapes, varying injection rates.  However, none of these changes have resulted in disks that simultaneously precess in the prograde and retrograde directions without inducing artificial means.  To induce both superhump modulations and both precessions, the run is stopped after prograde precession has begun, the disk is artificially rotated, and the run is restarted, as shown in some of these works. Although negative superhumps and retrograde precession are now present, the act to induce is artificial.  Hence we return to the root code to see if accretion disks can tilt naturally and precess in both directions. Results such as frequency values and amplitudes of signals can then be compared to results obtained from the control group, simulations using the same base code but artificially inducing a 5$^{o}$ disk tilt (Montgomery 2009a). 

The root code (Simpson 1995) uses the Lagrangian method of SPH, has a maximum of 25,000 particles, assumes an ideal gas law with a low adiabatic gamma, does not include radiation effects or magnetic fields.  The only unknown is the Shukura-Sunyaev (1973) $\alpha$ viscosity.  We use nearly the same input parameters in Montgomery (2012a), but this time we want to simulate an SU UMa system:  We assume secondary-to-primary mass ratio $q=M_{2}/M_{1}=0.24$, a similar mass ratio to that used in Wood et al. (2011); primary mass $M_{1}=0.82$ $M_{\odot}$; viscosity coefficients $\alpha=0.5$ and $\beta=0.5$; smoothing length $h=0.02$, and separation distance $d=0.9 R_{\odot}$.  In all simulations, the primary and secondary stars are treated as point masses, and gas particles are injected through the inner Lagrange point $L_{1}$ to build the disk from scratch.  

\section{Results and Analysis}
The simulation produces $\emph{early}$ positive superhumps ($P_{+}=1.075$ $P_{orb}$, where $P_{orb}$ is the orbital period, $P_{+}$ is the positive superhump period) and prograde precession once the superoutburst is initiated near orbit 65 (i.e., $\sim$5.7 days in V3444 Lyrae if the disk is built from scratch).  The outer annuli of the disk cyclically changes shape from circular to highly eccentric; two asymmetric spiral waves build and dissipate like those shown in Montgomery (2012b) and Whitehurst (1984, 1988) but do not appear to advance; and only the fundamental apsidal superhump frequency $\nu_{a}$ is present in the Fourier transform like that shown in Montgomery (2012b).  

Unlike other attempts to model V344 Lyrae (Wood et al. 2011), we continue the simulation beyond orbit 420 ($\sim$37 days in V344 Lyrae) as additional time is needed for settling of relatively colder particles to the disk midplane.  Until orbit $\sim$965 (i.e., $\sim$900 orbital periods or $\sim$85 days in V344 Lyrae after superoutburst initiation), only 2$\nu_{a}$ and 3$\nu_{a}$ higher order harmonics have since initiated,  and spiral like structures in outer annuli weakly advance (not shown). Near orbit $\sim$965, the disk begins to tilt and precess in the retrograde direction.  In addition to the already present apsidal mode frequencies, the fundamental nodal superhump $\nu_{n}$ and linear combination frequencies are now present (see Figure 1 and Table 1).  Two independent precessions are now occurring simultaneously in the same disk, prograde and retrograde.  

As shown in Figures 2 and 3, the disk is tilted and precessing in the retrograde direction as well as simultaneously cyclically changing shape from circular to highly eccentric and precessing in the prograde direction. Based on the observational data for V344 Lyrae (Wood et al. 2011), $\epsilon_{+}$/$\epsilon_{-}$ = 1.8 where ($\epsilon_{+}$ = $P_{+}/P_{orb}$ -1), ($\epsilon_{-}$ = 1 - $P_{-}/P_{orb}$), and $P_{-}$ is the negative superhump period.  The ratio $\epsilon_{+}$/$\epsilon_{-}$  is related to the rate of advance of the line of apsides relative to the rate of regression of the line of nodes.  The simulation does not produce the same fractional period excess $\epsilon_{+}$ (4.4\%, observed; 7.5\%, simulated) or fractional period deficit $\epsilon_{-}$ (2.5\%, observed; 3.9\%, simulated), but does produce nearly the same $\epsilon_{+}$/$\epsilon_{-}$ ratio.  To obtain more accurate individual  fractional excesses, better estimates of parameter(s) like stellar masses are needed.  To obtain a more precise ratio of the fractional periods, tidal torques on differing disk thickness along the rim during precession may need to be included in the definition, which is currently based on a uniform thickness disk that is solely precessing either in the prograde or retrograde direction.  

Figure 2 is in black and white to emphasize the location of the two spiral density waves in the outer-to-middle annuli relative to the location of the bright spot, the spot where the gas stream strikes the rim of the disk.  From time $t$=126.6 days to $t$=127.0 days in V344 Lyrae, the spiral waves advance in the prograde direction.  When the disk is highly eccentric at $t$=126.2 days and $t$=127.4 days, the density waves compress with the disk rim (i.e., the rim on the far side of the disk, furthest from the secondary star, compresses into the spiral wave on the far side of the disk; meanwhile, the spiral wave on the near side of the disk compresses into the rim of the disk on the near side; the compressions are sequential, not simultaneous).  When the disk is more circular at $t$=126.6 days and $t$=127.0 days, the density waves appear in the outer-to-middle disk annuli.  The two waves are asymmetric.  When compared with the control group of artificially tilted disks (Montgomery 2009b), the nodal superhump fundamental and higher order harmonic frequencies (see Table 1 for this simulations' values) agree within the uncertainties.  However, the amplitude of the signals do not agree, indicating that the disk tilt in this simulation is tilted less than four degrees.  To produce the stronger nodal superhump frequency in V344 Lyrae (Wood et al. 2011), the mass transfer rate may need to be increased to a rate more indicative of a system in outburst.  

Figure 3 shows yellow as a warmer emission color than orange.  The black color emphasizes relatively colder particles, which comprise the disk midplane; the two density waves; and the gas stream overflow.  The cold, dense disk midplane along the rim is warped as shown at $t$=126.6 days and $t$=127.4 days.  The disk is also slightly titled as shown at $t$=126.2 days and $t$=127.0 days, causing the gas stream to flow over the disk for $\sim$1/2 orbital period and under the disk form the latter $\sim$1/2 orbital period.  As suggested in the figure, the midplane may be warped enough that portions become part of the disk surface material.

\section{Discussion}
To generate a disk tilt, simulations in Montgomery (2012a) and this work use a continuous infall of equal size and equal mass particles, the path of which is in the orbital plane.  The infall strikes the disk at the bright spot and unequally overflows and underflows the disk rim, resulting in the disk's angular momentum $\bf{\emph{L}_{d}}$ becoming misaligned to the stellar host's rotational angular momentum and the system's orbital angular momentum $\bf{\emph{L}_{o}}$ yet the gas stream remains normal to $\bf{\emph{L}_{o}}$.  The differential flow paths over and under the disk may produce the disk tilt by the lift force (Montgomery \& Martin, 2010).  

Once the disk is tilted as a unit, the disk precesses in the retrograde direction by tidal torques from the secondary on the tilted disk as shown in Montgomery (2012a).  More of an undulating warp is generated in this simulation because the disk is simultaneously precessing in both the prograde and retrograde directions.  The changing thickness along the disk rim coupled with the continuous gas stream impact on the titled disk rim results in the discontinuous undulating warp. 

Early positive superhumps are generated in this simulation's artificial light curves, similar to those generated by the $q=0.25$ simulation in Wood et al. (2011), and thus we do not show these light curves in this work.  The observed light curves in V344 Lyrae show dominating positive superhumps followed by dominating negative superhumps. The artificial light curves generated in this work do show dominating early positive superhumps but do not generate prominent negative superhumps.  Instead, modulated superhumps are generated, the modulation being caused by the addition of  disk tilt and retrograde precession.  To generate dominating negative superhumps, the mass transfer rate may need to be increased to that of a system in outburst, a rate not simulated in this work.  This fine-tuning analysis is reserved for a later paper.  Also reserved for a later paper is harmonic synthesis of the simulated light curves and/or harmonic analysis of the observed and simulated light curves that produced Figure 1 and Table 1.  Modulated superhumps caused by simultaneous precessions are not discussed in Wood et al. (2011). 

Although the frequencies in Table 1 agree, within the uncertainties, with values reported in an artificially induced tilted disk simulation (Montgomery 2009a), the frequency amplitudes do not agree.  As the control group was artificially tilted five degrees, resulting in prominent negative superhumps  and large amplitude frequencies, the disk tilt in this simulation is less than four degrees.  Although a signal may be present in a periodogram that suggests the disk is precessing in the retrograde direction, the light curve may not show a prominent negative superhump as the disk is not tilted high enough.  

In this work, we show negative superhump frequencies and retrograde precession are possible in systems below the period gap (2-3 hours) such as in SU UMa systems like V344 Lyrae, V503 Cygni (Harvey et al., 1995), and SDSS J210014.12+004446.0 (Olech et al. 2009), all of which also show positive superhumps.  In Montgomery (2012a), we show negative superhump frequencies and retrograde precession are possible in systems above the period gap such as V378 Pegasi (Ringwald et al., 2012).  Thus negative superhumps and/or their signals should be ubiquitous above, below, and in the period gap like TT Ari.  Observers of CV DNs should thus be seeking these ubiquitous negative superhumps and/or frequencies to confirm this prediction.  

We note this simulation may not be over, and we report any additional cycles in a follow-up paper.  Also, we note that although the code is set to the parameters of an SU UMa system, many features shown in this work should also apply to other system types such as inner annuli of protoplanetary disks that may contain a planet in a gap.  

\section{Conclusions}
We find the following outcomes:
1) The observed light curves of V344 Lyrae shown in Wood et al. (2011) indicate cyclic times when positive superhumps dominate followed by times when negative superhumps dominate.  The simulation of V344 Lyrae in this work also produces dominating positive superhumps prior to the addition of the negative superhump frequencies, although the simulation does not produce dominating negative superhumps.  As such, the mass transfer rate may need to be modified in the simulation to that more of a system in outburst to induce the dominating negative superhumps, a test not included in this work. As suggested by this work, although a signal in a periodogram may indicate that the disk is precessing in the retrograde direction, the negative superhump in the observed light curve may not be present.
2) As shown in Montgomery (2012a) and this work, neither a superoutburst (which is associated with positive superhumps) nor an outburst is required to induce disk tilt and retrograde precession (which are associated with negative superhumps).
3) The simulation of V344 Lyrae in this work produces simultaneous disk prograde and retrograde precession, which is not discussed in Wood et al. (2011). The observations (Wood et al. 2011) do not find a significant signal at the retrograde precession period \( P_{p-}^{-1} = P_{-}^{-1} - P_{orb}^{-1} \sim \) 3.6 days.  Similarly, no significant simulated signal at the retrograde precessional period ($\sim$2.2 days) is found in Figure 1. 
4) Although individual observed and simulated fractional periods $P_{-}$ and $P_{+}$ do not agree, nearly the same ratio of fractional period excess $\epsilon_{+}$ to fractional period deficit $\epsilon_{-}$ is obtained. This result implies that stellar masses are not precise but are accurate in the simulation.  
5) The frequencies in Table 1 agree, within the uncertainties, with values reported in an artificially induced tilted disk simulation that is not precessing in the prograde direction (Montgomery 2009a).  The frequency amplitudes do not agree, however, suggesting that the disk is not tilted as high.     
6) We show in this work that negative superhump frequencies and retrograde precession are possible in systems below the period gap (2-3 hours).  In Montgomery (2012a), we show negative superhump frequencies and retrograde precession are possible in systems above the period gap.  Thus negative superhumps and/or their signals should be ubiquitous above and below the period gap.   
 
\section*{Acknowledgments}
This research was supported in part by NASA through the American Astronomical Society's Small Research Grant Program.  We would like to thank NSF UCF YES recipient and undergraduate student, Nicholas Howell, who helped analyze disk effects when the first higher order harmonic is present in tilted disks that only precess in the retrograde direction.  We also kindly thank the reviewer whose careful analysis and comments improved this work. 
%\begin{thebibliography}{99}
%\bibitem{b1} Harvey, D., Skillman, D.R., Patterson, J., \& Ringwald, F.A., 1995, PASP, 107, 551
%\bibitem{b2} Horne, K., 1984, Nature, 312, 348
%\bibitem{b3} Lasota, J.-P., 2001, New Astron. Rev., 45, 449
%\bibitem{b4} Montgomery, M.M., 2009a, MNRAS, 394, 1897
%\bibitem{b5} Montgomery, M.M., 2009b, ApJ, 705, 603 
%\bibitem{b6} Montgomery, M.M., 2012, ApJ, 745, L25
%\bibitem{b7} Montgomery, M.M., 2012b, in The Golden Age of Cataclysmic Variables and Related Objects, F. Giovannelli \& L. Sabau-Graziati (eds.), Mem. SAIt. 83 N.2 
%\bibitem{b8} Montgomery, M.M., 2004, PhD Thesis, Florida Institute of Technology
%\bibitem{b9} Montgomery, M.M. \& Bisikalo, D.V., 2010, MNRAS, 405, 1397
%\bibitem{b10} Montgomery, M.M. \& Martin, E.L., 2010, ApJ, 722, 989
%\bibitem{b11} Olech, A., Rutkowski, A., \& Schwarzenberg-Czerny, A., 2009, MNRAS, 399, 465
%\bibitem{b12} Pringle, J.E., 1996, 281, 357
%\bibitem{b13} Ringwald, F.A., Velasco, K., Roveto, J.J., \& Meyers, M.E., 2012, NewA, 17, 433
%\bibitem{b14} Shakura S.. \& Sunyaev R.A., 1973, A\&A, 149, 135
%\bibitem{b15} Skillman D.R., 1998, ApJ, 503, L67
%\bibitem{b16} Simpson, J.C., 1995, ApJ, 448, 822
%\bibitem{b17} Whitehurst, R. 1988, MNRAS, 232, 35
%\bibitem{b18} Whitehurst, R., 1994, MNRAS, 266, 35
%\bibitem{b19} Wood, M.A., Montgomery, M.M., \& Simpson, J.C., 2000, ApJ, 535, L39
%\bibitem{b20} Wood, M.A., Still, M.D., Howell, S.B., Cannizzo, J.K., Smale, J.P., 2011, ApJ, 741, 105
%\bibitem{b21} Wood M.A., Thomas D., Simpson J.C., 2009, 398, 2110
%\end{thebibliography}

\section{References}
Harvey, D., Skillman, D.R., Patterson, J., \& Ringwald, F.A., 1995, PASP, 107, 551
Horne, K., 1984, Nature, 312, 348
Lasota, J.-P., 2001, New Astron. Rev., 45, 449
Montgomery, M.M., 2009a, MNRAS, 394, 1897
Montgomery, M.M., 2009b, ApJ, 705, 603 
Montgomery, M.M., 2012, ApJ, 745, L25
Montgomery, M.M., 2012b, in The Golden Age of Cataclysmic Variables and Related Objects, F. Giovannelli \& L. Sabau-Graziati (eds.), Mem. SAIt. 83 N.2 
Montgomery, M.M., 2004, PhD Thesis, Florida Institute of Technology
Montgomery, M.M. \& Bisikalo, D.V., 2010, MNRAS, 405, 1397
Montgomery, M.M. \& Martin, E.L., 2010, ApJ, 722, 989
Olech, A., Rutkowski, A., \& Schwarzenberg-Czerny, A., 2009, MNRAS, 399, 465
Pringle, J.E., 1996, 281, 357
Ringwald, F.A., Velasco, K., Roveto, J.J., \& Meyers, M.E., 2012, NewA, 17, 433
Shakura S.. \& Sunyaev R.A., 1973, A\&A, 149, 135
Skillman D.R., 1998, ApJ, 503, L67
Simpson, J.C., 1995, ApJ, 448, 822
Whitehurst, R. 1988, MNRAS, 232, 35
Whitehurst, R., 1994, MNRAS, 266, 35
Wood, M.A., Montgomery, M.M., \& Simpson, J.C., 2000, ApJ, 535, L39
Wood, M.A., Still, M.D., Howell, S.B., Cannizzo, J.K., Smale, J.P., 2011, ApJ, 741, 105
Wood M.A., Thomas D., Simpson J.C., 2009, 398, 2110

\newpage

 %%FIGURE 1 %%%%%%%%%%%%%%%%%%%%%%%%
\begin{figure}
\includegraphics[scale=0.9]{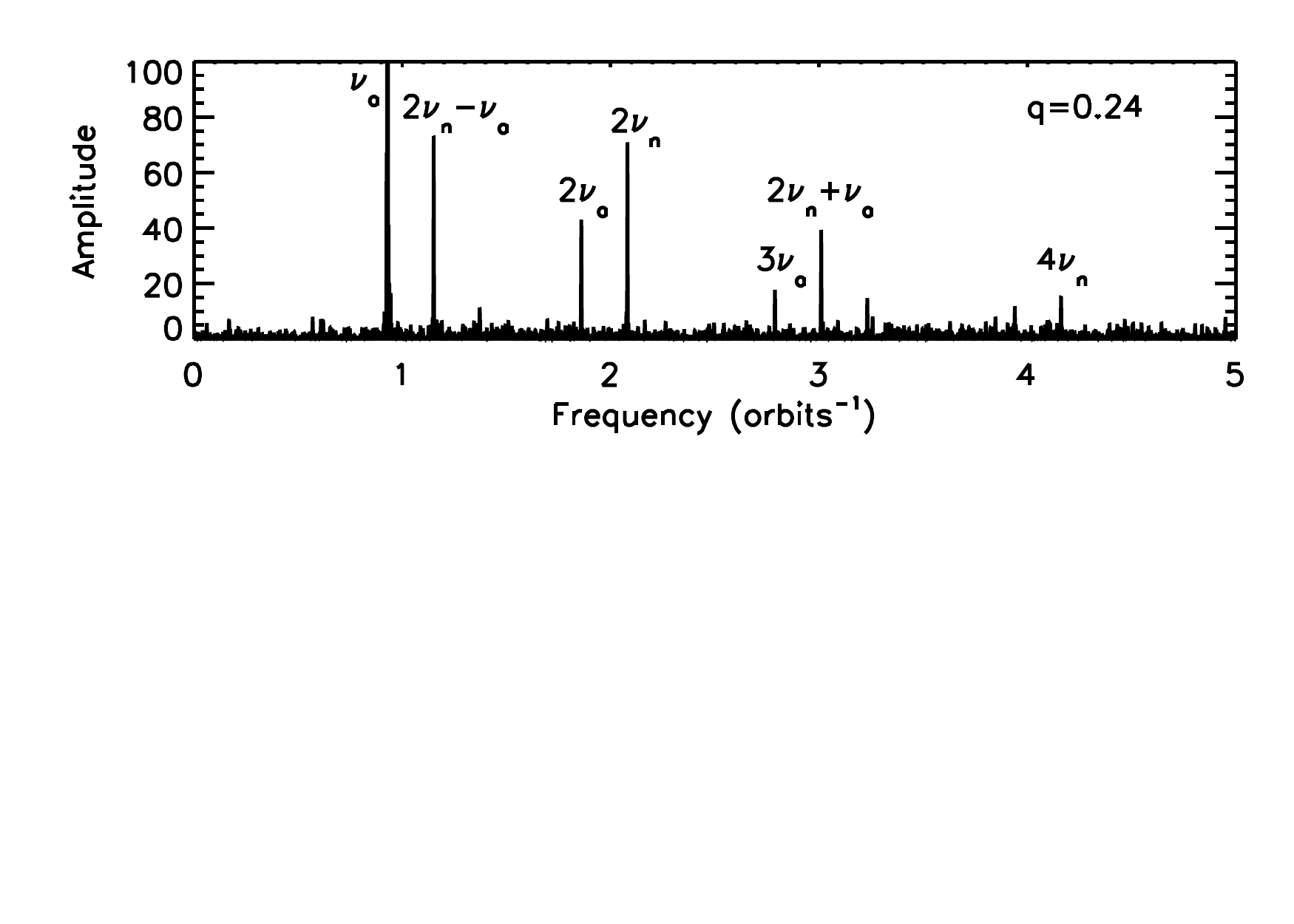}
\caption{Fourier amplitude spectrum of orbits 1200-1500 where one orbit equals one orbital period.  Mode frequencies are defined and listed in Table 1.  
}
\label{Figure 1.}
\end{figure}
%%%%%%%%%%%%%%%%%%%%%%%%%%%%%%%%%%%

\newpage
 %%FIGURE 2 %%%%%%%%%%%%%%%%%%%%%%%%
\begin{figure}
\includegraphics[scale=0.6]{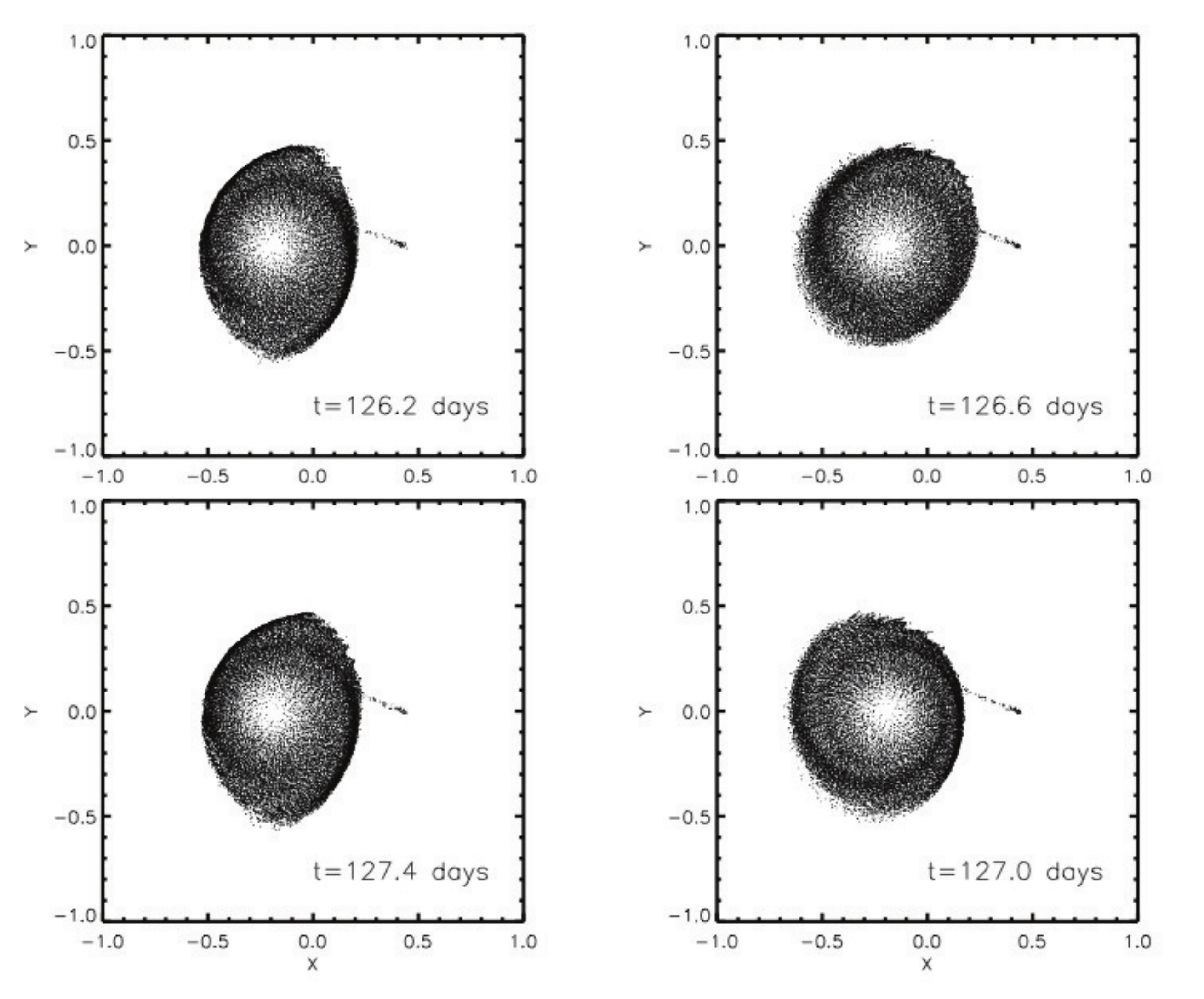}
\caption{Numerical simulation of an accretion disk around a white dwarf star (not shown) that is accreting material from a close secondary star (not shown), similar to the accretion disk around V344 Lyrae. The four panels are face-on snapshots in time, taken at the same phase in the orbit, showing gaseous material injected from $L_{1}$ to the disk rim.  Twenty orbital periods or $\sim$1.2 days in V344 Lyrae are shown.  As time progresses, the outer annuli of the disk cyclically change shape from highly eccentric to circular to highly eccentric. During this disk shape-change cycle, two asymmetric spiral density waves in the outer-to-middle annuli compress with the disk rim and advance $\sim$180$^{\circ}$ in the prograde direction.  Two shape-change cycles are needed to advance the spirals arms once.
 }
\label{Figure 2.}
\end{figure}
%%%%%%%%%%%%%%%%%%%%%%%%%%%%%%%%%%%

\newpage
 %%FIGURE 3 %%%%%%%%%%%%%%%%%%%%%%%%
\begin{figure}
%\rotatebox{90}{\includegraphics[scale=0.5]{Fig3.pdf}}
%\includegraphics[scale=0.9]{Fig3.eps}
\includegraphics[scale=0.9]{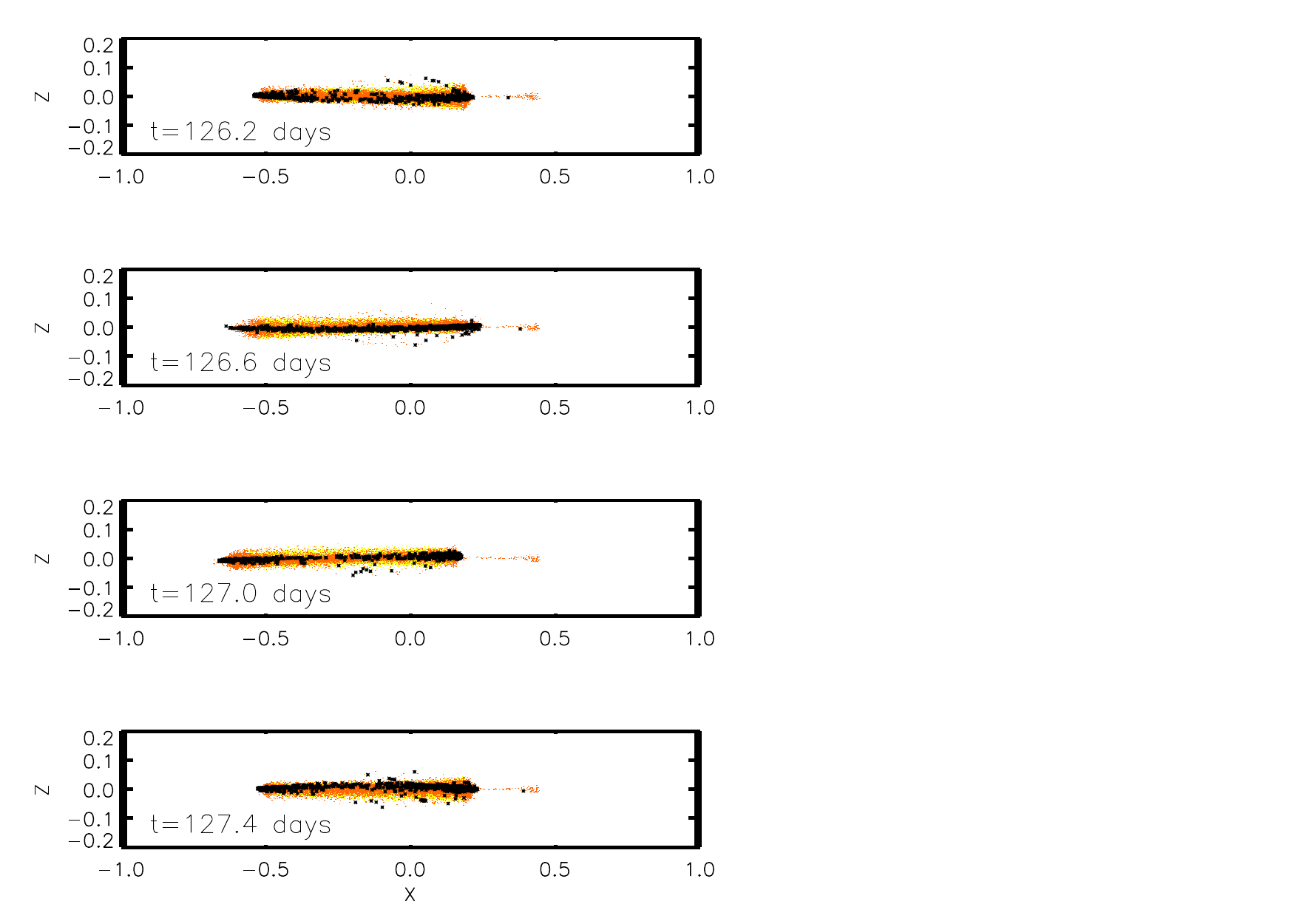}
\caption{Edge-on snapshots are shown at times in the simulation described in Figure 1.  Yellow colored gas particles are warmer than orange colored gas particles.  Black colored gas particles highlight relatively colder particles, which comprise the dense midplane; the gas stream overflow; and the two spiral arms.  As shown, the disk is now tilted in addition to changing shape.  Because the tilted disk is precessing in the retrograde direction, the location on the disk rim where the gas stream transitions from flowing over the disk to under the disk (i.e., bright spot) also precesses in the retrograde direction.  The net result is that the disk appears to wobble backwards as a unit, as shown, and the gas stream appears to flow over the disk ($t$=126.2 days) instead of under ($t$=126.6 days) when at the same phase in the orbit.  As shown, the cold, dense disk midplane is warped. }
\label{Figure 3.}
\end{figure}
%%%%%%%%%%%%%%%%%%%%%%%%%%%%%%%%%%%

\newpage
%%%%%%%%%%%%%%%%%TABLE 1 Class I and Class II Stellar Values%%%%%%%%%
\begin{table*}
 \centering
 \begin{minipage}{200mm}
  \caption{Apsidal $\nu_{a}$ and Nodal $\nu_{n}$ Mode Frequencies }
  \begin{tabular}{@{}lc@{}}
Mode     \& Frequency ($\pm$0.003 orbit$^{-1}$)  \\
 \hline
$\nu_{a}$  &  0.930\\
2$\nu_{a}$  &  1.860\\
3$\nu_{a}$  &  2.788 \\
2$\nu_{n}$* &  2.080 \\
4$\nu_{n}$  &   4.161\\
2$\nu_{n}$ - $\nu_{a}$ & 1.151 \\
2$\nu_{n}$ + $\nu_{a}$ &  3.010\\
\hline
\end{tabular}
\end{minipage}
*Nodal mode frequencies are negative; absolute values are shown; one orbit is one orbital period.
\end{table*}
%%%%%%%%%%%%%%%%%%%%%%%%%%%%%%%%%%%%%%%%%%%%%%%%%%%%%

\newpage

\end{document}